\documentclass[letterpaper,10pt,final,compsoc]{IEEEtran}
\usepackage[
left=1in,
right=1in,
top=1in,
bottom=2.86cm, 
columnsep=.81cm
]{geometry}
\usepackage[T1]{fontenc}
\usepackage[utf8]{inputenc}
\usepackage{mathptmx}
\usepackage{amsthm}
\usepackage{enumitem}
\makeatletter
\let\NAT@parse\undefined
\makeatother
\usepackage[bookmarks=false,colorlinks,urlcolor=blue,%
linkcolor=magenta,citecolor=red,linktocpage=true]{hyperref}
\usepackage{xspace}
\usepackage{graphicx}
\usepackage{inconsolata}

\usepackage[textsize=footnotesize,linecolor=teal,%
bordercolor=teal,backgroundcolor=lime!25]{todonotes} 
\newcommand{\sm}{\textsf{SMAll}\xspace}

\makeatletter
\def\@IEEEsectpunct{.\ \,}
\def\paragraph{\@startsection{paragraph}{4}{\z@}{1.5ex plus 1.5ex minus 0.5ex}%
{0ex}{\normalfont\normalsize\itshape}*}
\makeatother

\title{\LARGE Insider Threats in Emerging Mobility-as-a-Service Scenarios}

\author{
	\IEEEauthorblockN{Franco Callegati$^1$,
	Saverio Giallorenzo$^{2,3}$,
	Andrea Melis$^2$, and
	Marco Prandini$^2$
	}
	\IEEEauthorblockA{
		\\[.5em]
		{\small Dep. of \{
			$^{1}$Engineering and Information Systems, 
			$^{2}$Computer Science and Engineering
		\}, University of Bologna, $^{3}$INRIA%
	}}
}

\begin{document}

\maketitle
\thispagestyle{empty}
\pagestyle{empty}

\begin{abstract}
\normalfont\itshape
Mobility as a Service (MaaS) applies the everything-as-a-service paradigm of
Cloud Computing to transportation: a MaaS provider offers to its users the dynamic composition of solutions of different travel agencies into a single, consistent interface.

Traditionally, transits and data on mobility belong to a scattered plethora of
operators.
Thus, we argue that the economic model of MaaS is that of federations of
providers, each trading its resources to coordinate multi-modal solutions for
mobility.
Such flexibility comes with many security and privacy concerns, of
which insider threat is one of the most prominent.
In this paper, we follow a tiered structure --- from individual operators to
markets of federated MaaS providers --- to classify the potential
threats of each tier and propose the appropriate countermeasures, in an effort
to mitigate the problems.
\end{abstract}

\section{Introduction} 
\label{sec:introduction}

The term Cloud Computing denotes a dynamic infrastructure where users access
services without regard to where the services are hosted~\cite{Buyya2009599}.
The concept of Mobility as a Service (MaaS)~\cite{MaaS} takes inspiration from
such a model and brings it into the context of transportation.
In Cloud Computing, the architecture that runs the services is dynamic and
transparent to users. Likewise, MaaS hides a dynamic composition of solutions
provided by different travel agencies behind a consistent interface. Hence, MaaS
users experience traveling as provided by a single agency.

Due to regulatory and logistic issues, mobility resources are administrated and
owned by a scattered plethora of mobility operators (traditional travel
agencies and providers of data for mobility). Thus, we argue that the leading
economic model of MaaS markets is that of federations of mobility operators,
each trading its resources.
In such a federated market, operators can dynamically partner with each other,
still preserving their individual autonomy and without the need for a
centralized regulation authority.
On these premises, we are currently developing a Service-Oriented platform,
called \emph{Smart Mobility for
All}\footnote{\url{https://github.com/small-dev/SMAll.Wiki/wiki}} (\sm), built
on the concept of federated Cloud
Computing~\cite{rochwerger2009reservoir,buyya2010intercloud} and purposed to
support liquid markets for transportation.

During the development of \sm and through the collaboration with our industrial
partners (public administrations, local travel agencies, etc.), we identified
and analyzed many security issues spanning from a single operator to a
federation of operators.
In this context, we deem malicious insider activity one of the most prominent,
spanning from standard threats against cloud installations~\cite{Kandias2013}
to insider issues specific to the contexts of mobility and of markets of
services.

\begin{figure*}[tb]
  \centering
  \includegraphics[width=\textwidth]{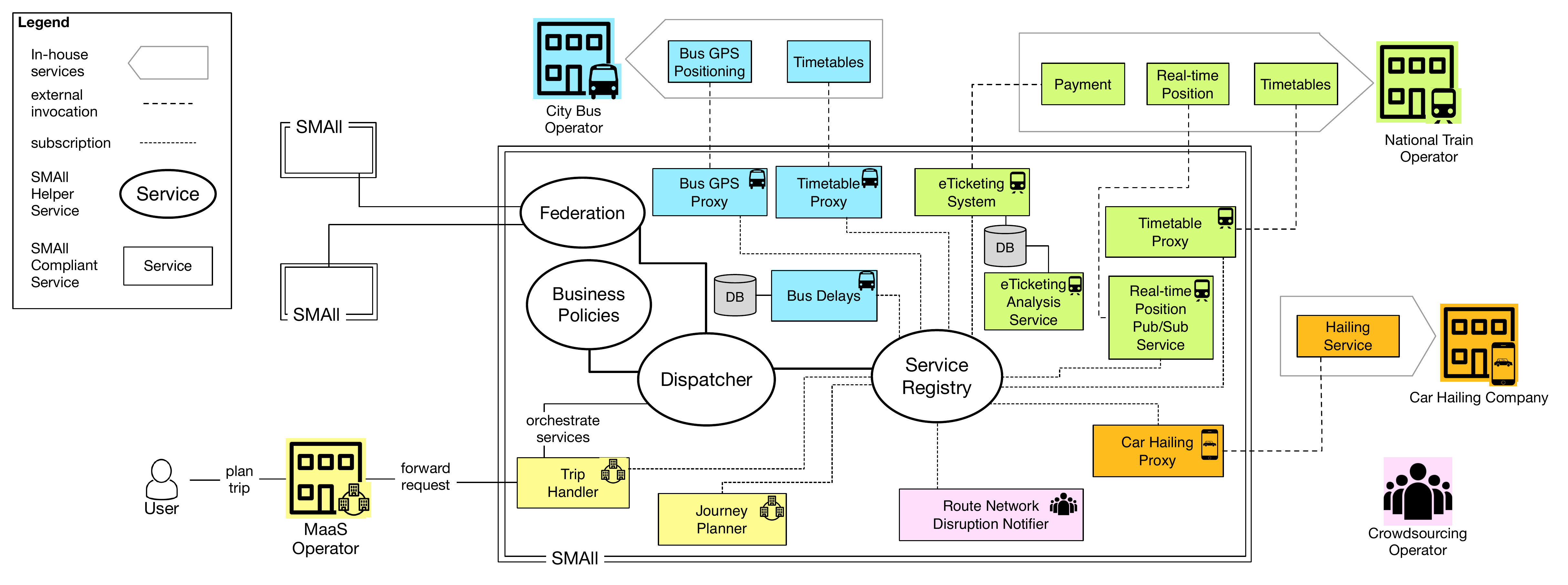}
  \caption{Example of the \sm architecture.}
  \label{fig:small_architecture}
\end{figure*}

\noindent\textbf{Motivation.} Fig.~\ref{fig:small_architecture} depicts a
cross section of an instantiation of \sm, where the colored entities outside
of the boundaries of \sm (bordered with double lines) are public
transportation agencies, private companies, on-line communities, and MaaS
operators.
  
Even when considered in isolation, the agents in the platform already entail
well-known threats due to insider activity. For example, the City Bus Operator
represents a threat to the privacy of drivers since GPS positioning can reveal
sensitive information on their conduct, which is forbidden under some
legislation; however, also drivers represent an insider threat to the Bus
Operator: they can disable the GPS device on their vehicles, compromising the
reliability of the GPS positioning system and that of the other services that
depend on it\footnote{The issues are far from being just speculative, as we
actually encountered them collaborating with one of our industrial partners.}
(e.g., the Bus Delays service that estimates bus arrivals based on vehicle GPS
positions). Finally employees can manipulate the services and their data,
damaging the company by extracting restricted information or causing outages.

Broadening our scope to federated interactions, we focus on the MaaS Operator
in Fig.~\ref{fig:small_architecture} that, for example, deploys a Journey
Planner service for providing dynamic multi-modal trips to users. The
service orchestrates other federated services in \sm: it uses information on
scheduling, availability, disruptions, and the position of buses, trains, and
on-demand cars. 

As expected, the threats highlighted for single operators surface (and possibly
combine) to higher-level federated scenarios. Consider the case in which the
City Bus Operator allows the MaaS Operator to access the Bus GPS Proxy service.
With the raw data on the real-time position of buses, the MaaS Operator can
undertake many malicious activities to the detriment of the Bus Operator, e.g.,
passing relevant information to competitors of the Bus Operator.
Another important threat comes from the extraction of sensitive information
from aggregated/anonymized data. Aware of the threat posed by the Bus GPS Proxy
service, the Bus Operator markets only its Bus Delays service. However, also
aggregated data like the temporal approximation of the arrival of buses might
let the MaaS Operator extract~\cite{zhou2012long} the actual position of
vehicles (possibly optimizing the accuracy of the extraction~\cite{mirri2016}).

\noindent\textbf{Contribution.} As exemplified, in the context of MaaS
operators, the definition of what an insider is can assume subtle nuances
depending on the considered scenario. In this work, guided by our experience
with the development of \sm, we describe the security issues concerning
insiders within such a federated market of services.
We develop our treatment following a tiered view of MaaS markets, called the
MaaS Stack~\cite{maaS_stack}, summarized in
\S~\ref{sec:mobility_as_a_service}. In \S~\ref{sec:insider_theats}, we consider
each tier of the MaaS Stack, we define what an insider is for each of them, we
analyze the related threats, and we describe the possible countermeasures.

\section{The MaaS Stack: An Overview} 
\label{sec:mobility_as_a_service}
\begin{figure*}[tb]
  \centering
  \includegraphics[width=\textwidth]{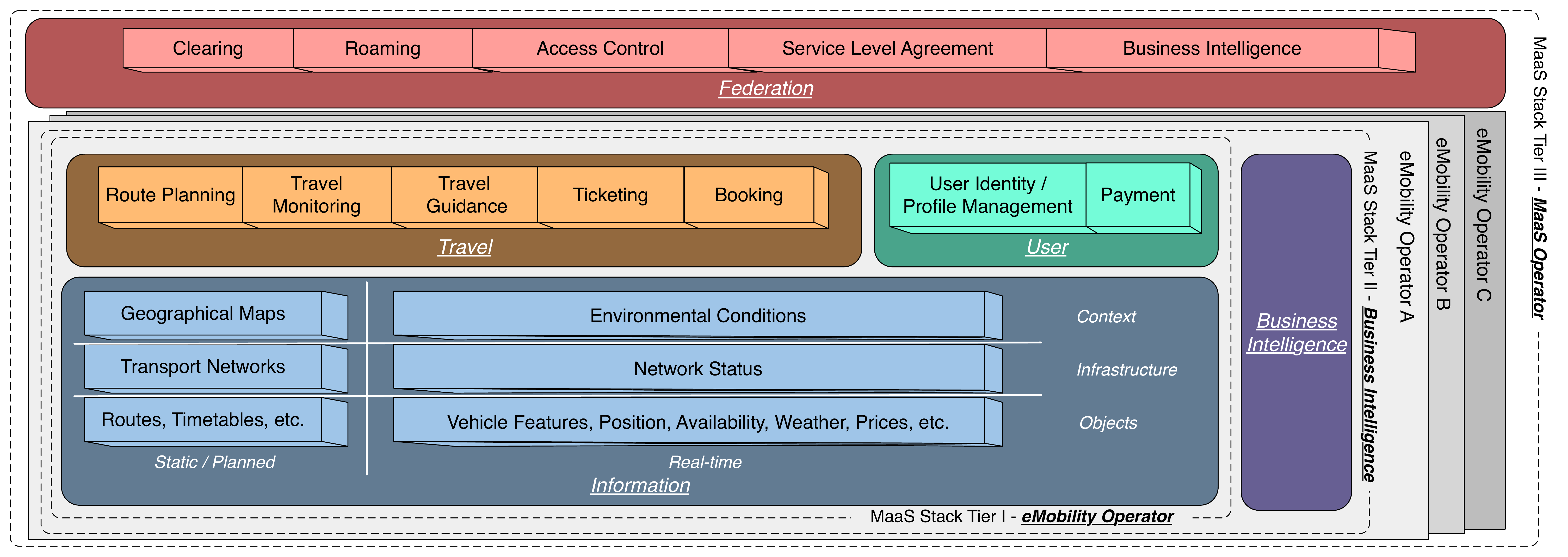}
  \caption{The MaaS Stack}
  \label{fig:MaaS_Stack}
\end{figure*}

In this section we briefly overview the MaaS Stack (Fig.~\ref{fig:MaaS_Stack}),
a structured view that we assembled to guide the development of \sm. In
\S~\ref{sec:insider_theats} we use the MaaS Stack to analyze the insider
threats of each tier.

\noindent\textbf{Tier I | eMobility Operators.}
The first tier of the MaaS Stack is that
of \emph{eMobility Operators}. An eMobility Operator is an entity that owns,
administrates, and exposes software functionalities regarding mobility,
provided in a machine-readable form. In tier I of the MaaS Stack, eMobility
Operators are considered in isolation (i.e., not using and integrating the
services of other operators).
For example, the National Train Operator represented in 
Fig.~\ref{fig:small_architecture}
is an eMobility Operator that owns services for purchasing tickets, accessing
timetables, and receiving real-time position of vehicles.

\noindent\textbf{Tier II | Business Intelligence.} The second tier of the MaaS
Stack still focuses on single eMobility operators but it enriches the taxonomy
of services with the category of \emph{Business Intelligence}. These services
are not meant for users but for eMobility operators; they span
over first-tier services by monitoring and analyzing their usages. Business
Intelligence services provide insight on the performances of eMobility
operators. For example, the eTicketing Analysis Service
(Fig.\ref{fig:small_architecture}) of the Train Operator can suggest to the
latter new pricing policies as well as reporting rarely used routes that could
be merged/discarded.

\noindent\textbf{Tier III | MaaS Operators.}
The last tier of the MaaS Stack is that of MaaS Operators, i.e., eMobility
operators that federate and integrate their services with those of other
eMobility operators. Each MaaS operator provides to its users information and
transit services of other operators as its own. The principle resembles that of
``roaming'' in GSM phone networks~\cite{mouly1992gsm}, where users connect
through the services of another phone company when traveling outside the
geographical coverage area of the home network.
The MaaS Operator represented in Fig.~\ref{fig:small_architecture} can 
federate with the
National Train Operator and the City Bus Operator and it can offer multi-modal
journeys that span different means of transportation and have nation- to
city-wide scopes.
This example introduces the last fundamental element of the third tier of the
MaaS Stack: \emph{Clearing} services to account for federated usages and
compensate operators according to the established Business Policies.

%
To support the mentioned features in \sm, we are currently developing and
integrating components to deploy services, to support the definition and
enforcement of business and clearing policies, and to federate many instances
of the platform.
During the development, we recognized and investigated 
security issues derived from the openness of our federated platform.
In the next section, we consider each tier of the MaaS Stack, we define what
an insider is for each of them, we analyze the related threats, and we describe
the possible amendments to counteract them.

\section{Insider Threats in MaaS Scenarios} 
\label{sec:insider_threats}
\label{sec:insider_theats}

Statistically, insider threats are one of the most expensive security issues
for business companies~\cite{mun2008yet}. One prominent reason of these
expensive outcomes is that companies did not foresee all possible malicious
insider activities~\cite{Stavrou2014}. Indeed, the problem is not the lack of
proper countermeasures as much as the difficulty of identifying a malicious
insider in the first place. Literature abounds with guidelines and principles
aimed at providing general descriptions of the context and the identity of the
insiders \cite{flynn2014cloud,claycomb2012insider}. However, experts agree
that the strong contextual variance of threats~\cite{schneier2011secrets}
makes providing a general yet precise identification of all possible insiders
difficult.

Thus, we deem useful to share the experience we gained in the context of
services for mobility (both at software and physical level). Moreover, our
background on the development of \sm provides insights on the possible threats
deriving from federated cloud architectures, built for deploying, publishing,
and trading services. Federated clouds have been already analyzed in
literature~\cite{Kandias2013,nostro2014insider}, however we deem important to
include the related threats in the frame of the emerging Mobility-as-a-Service
scenario.

\begin{figure*}[tb]
	\centering
	\includegraphics[width=\textwidth]{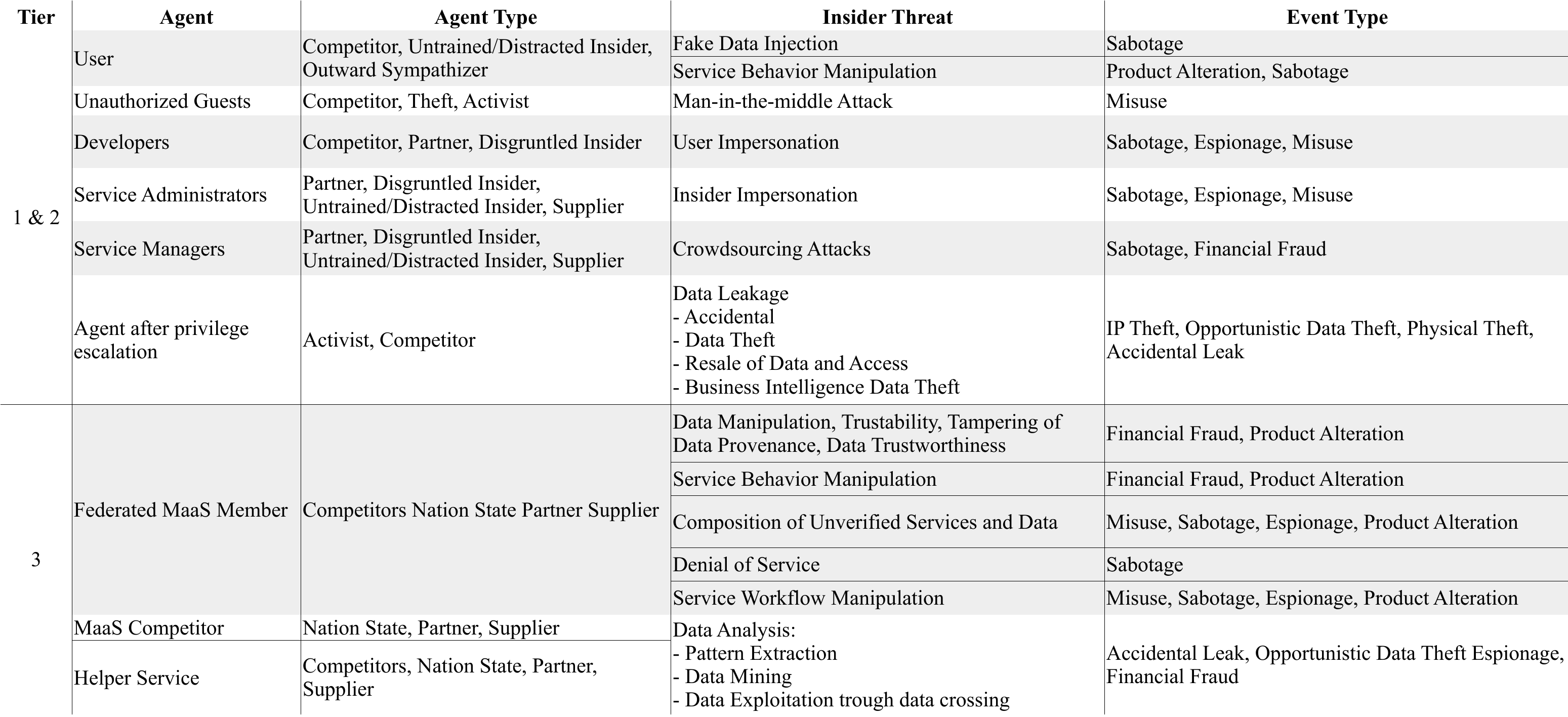}
	\caption{Summary table comparing the tiers of the MaaS Stack to the related insider threats.}
	\label{fig:summary_table}
\end{figure*}

In \S~\ref{sub:maas_stack_tier_i}--\ref{sub:maas_stack_tier_iii}, we
illustrate, for each tier of the MaaS Stack
(cf.~Fig.~\ref{sec:mobility_as_a_service}), the insiders, the related attacks,
and the possible countermeasures, as found in the state of the art and
implemented in \sm. In Fig.~\ref{fig:summary_table} we report a table that
summarizes our findings. Agents and threats are classified according
to the categories identified by Casey in \cite{casey} and the CERT technical
report \cite{silowash2012common}.

We dedicate the last paragraph of this Section to a brief description of the
methodology we followed to recognize the threats and the respective
countermeasures.

\paragraph{Methodology}
As mentioned, adopting a precise definition of insider may hinder the
identification of threats specific to particular contexts. Therefore, in our
investigation, we prefer to look at insiders from a general point of view~\cite{bishop2005position}:
\begin{quote}\textit{ 
A trusted entity that is given the power to violate one or more rules in a
given security policy [\textellipsis] the insider threat occurs when a trusted
entity abuses that power.}
\end{quote} 
This definition hints that an insider is determined by the role played as
member of a system and related to the deployed control rules and the pursuable
malicious goal(s).
In our context, the most classic scenario is one where the insider is
within the service of the victim, e.g., a programmer that manipulates
the behavior and the data of a service.
However, orchestrations spanning many providers, hallmark of the \sm platform,
lead to subtle yet relevant threats. Consider the case of federated partners.
On one hand, the provider of a service exposes itself to threats posed by
members that use its service --- security issues span from misuse of
information extracted from the service to over-usages that entail unforeseen
costs or outages --- on the other hand, an agent that orchestrates services of
other partners is a man-in-the-middle able to leak private information,
counterfeit data or use its vantage point to extract strategic patterns from
partners.

Regarding countermeasures, we structured our analysis of the possible
alternatives following the review compiled by Hunker and
Probst~\cite{hunker2011insiders}, encompassing the three approaches:
\emph{i}) \emph{Prevention}, comprising the definition of strong access control
	rules, data management systems (including data masking and data camouflage),
	and mechanisms to guarantee data provenance and data trustworthiness;
\emph{ii}) \emph{Detection}, that usually goes hand-in-hand with dissuasion mechanisms
	such as techniques of data management and service invocation that make abuses
	extremely expensive in terms of computing power;
\emph{iii}) \emph{Mitigation}, that exploits auditing and monitoring techniques, based on
	machine learning, to automatically identify and react to insiders.

\subsection{MaaS Stack | Tier I} 
\label{sub:maas_stack_tier_i}

As reported in \S~\ref{sec:mobility_as_a_service}, the first tier of the MaaS
Stack focuses on single eMobility operators and categorizes their services.
In this tier, the ecosystem of services has a flat structure and all members
play the same role of providers, without any interaction between each other.
Here, insiders can be pinpointed within two types: \emph{i})
\emph{users} authorized to interact with services and \emph{ii}) the
\emph{managers} (also seen as owners) of the services. In the reminder, we call
Users the members of the first type and Managers the members of the second one.
The distinction between the two types is trivial: while Users have limited
access to data and functionalities of a service, Managers can have full or
partial control (depending on the responsibility level) over the life-cycle of
the service and its resources.

Users allowed to interact with \sm services can basically pose two types of
threats:  \emph{i}) perform fake data injection (for crowdsourcing-based
services) and \emph{ii}) sharing the access to the services or to the
respective data.
Users can also exploit vulnerabilities to acquire Manager privileges
(configuring an \emph{Insider Impersonation} threat). However, we do not
include a discussion on these kind of attacks as they coincide with those
described for Managers.
Regarding Managers, their main threats comprise:
\begin{itemize}
\item manipulation of the behavior of a service, i.e., the computations done by
a service;
\item manipulation of the workflow among services, i.e., the flow of
	information among services;
\item stealing data, metadata, and performing malicious analyzes;
\item exposing sensitive information.
\end{itemize}

Following the first tier of the MaaS Stack, we describe the possible insider
attacks of each category of services.

\subsubsection{Information} 
\label{ssub:information}

The category of Information spans from basic services that publish raw
data (e.g., timetables or the position of vehicles) to higher-level services
that elaborate raw data to extract new information (e.g., the expected delay
of buses whose calculation requires the position of a vehicle and its scheduled
plan).
Notably, since Information services orchestrate other services to calculate and
publish these refined data, they are subject to \emph{Service Workflow
Manipulation} and \emph{Composition of Unverified Services and Data} threats.
We omit to present these issues in this Section and refer the discussion to
\S~\ref{ssub:composition}.

\paragraph{Data Leakage}
Data leakage is the accidental distribution of private or sensitive data to
unauthorized entities~\cite{shabtai2012survey}. In \sm, both Users and Managers
can cause data leakage. Users can share data to other, non-authorized Users.
Similarly, Users can also share their access to services, which could lead to
data leakage but also to other type of threats like \emph{User Impersonation}.
As expected, data leakage becomes even more serious when considered for Managers
that can share or steal sources unreachable by users.

\paragraph{Countermeasures}
Data leakage poses a serious issue in open networks where the transition of
data is not regulated nor monitored in their path. In these regards, \sm holds
a privileged position. In fact, all communications among the services in the
platform happen through the Dispatcher (cf. Fig~\ref{fig:small_architecture}),
which can log the quality and quantity of information required by all Users.
Obviously, this guarantee ceases when data exits the platform. The same
tracing system applies also to Managers.

\paragraph{Crowdsourcing Attacks}
Users can perform insider attacks on crowdsourcing services. These services
handle data streamed from sensors and devices or through direct signaling of
the users. An example is a crowdsourcing service where users can report
architectural limits for people with disabilities~\cite{mirri3_2016}. In this
case, insiders can feed the service with fabricated data to alter the normal
behavior of services, e.g., by directing users through specific pathways.

\paragraph{Countermeasures}
For the sake of completeness and clarity, let us start from the literature
regarding ``classic'' threat scenarios. Cho et al.~\cite{insider} examined how
insider attacks can exploit security holes in a trusted network of sensor
nodes. This work is of interest for our platform because it shows how even
trust-based approaches, in architectures that have to unify many nodes, are
not guaranteed to prevent attacks.

In~\cite{gertz2007handbook}, the authors described how access control policies
for a database management system can be exploited by insiders when the control
restrictions to be enforced may come from different authorities. Shatnawi et
al.~\cite{shatnawi2011detection} made a similar analysis but based on the
detection of malicious usage of a data source, which is equivalent to our case
of a malicious influence of data source services exposed by the \sm platform.

An interesting work that can be applied to our architecture
is~\cite{spitzner2003honeypots}. Here the authors implemented a pool of
honeypots to catch insiders. A honeypot is an information system resource
whose value lies in unauthorized or illicit use of that resource. The high
flexibility of honeypots --- able to play a huge variety of \sm-compliant
services --- is essential to make insiders expose themselves. Another useful
method that can be easily built within \sm is a reporting system for
crowdsensing and crowdsourced data, implemented in~\cite{mirri2_2016}. The
reporting system is based on the mapping of what the authors called Point of
Interest (POI). Each POI and its related data can be added to the system by
means of one or more reports. Reports are classified in three different source
classes, accordingly to the reputation of the user that collects the data.

\paragraph{Service Behavior and Data Manipulation}
As expected, insider threats posed by Managers constitute a more complex
scenario.
This type of insiders can access and modify the raw data of services as well as
manipulating their logic to present altered results.
Notably, since in our context the physical world mixes with that of software
services, we extend the role of Managers not only to the developers that can
modify the actual code of the service but also to conductors and other
operators: agents that can access and manipulate the physical devices that feed
the services.

The manipulation of these services can have many purposes from the point of
view of an insider. For example, during the development of \sm we interacted
with many industrial partners, among which there were some public
transportation companies that provided real-time positioning of their
vehicles. However, some of these companies did not report the actual position
of buses and instead published fake positions to mirror the exact planned
schedule. In another case the service worked intermittently. In the first
case, the company provided fake data to protect itself against possible
penalties due to delays, in the second case the positioning service went down
for certain rides due to drivers that disabled the in-vehicle positioning
devices either for fraudulent purposes (to avoid being scrutinized) or even
for shallow reasons such as to disable annoying automatic voice announcements.

\paragraph{Countermeasures}
Interesting works tackle the issue of how to predict insiders activities. Ho
et.~al.~\cite{ho2016demystifying} implemented a detection mechanism for single
users based on analyzes of changes on the writing style of the user after an
attack occurred, using machine learning algorithms. Althebyan
\cite{althebyan2008design} implemented a prediction model based on graph theory
approaches, to push alert once a detection risk mechanism finds that users are
performing actions that might lead to compromise the system services.

Studies also exist aimed at discovering malicious command execution. Among the
most relevant works, Kamra et al.~\cite{kamra2008detecting} and Mathew et
al.~\cite{mathew2010data} focus on the analysis of anomalous commands executed
on databases. In particular, they proposed a syntax analysis system to detect
anomalous queries; the former analyzed the submitted SQL queries, while the
latter focused on data retrieved from queries. Doss and
Tejay~\cite{doss2009developing} conducted a similar investigation as a field
study within an enterprise, where analysts were monitored while performing
their jobs.
Again, these results can be readily  applied in our architecture, especially
considering that tier I services will in any case be monitored by probes needed
to build Business Intelligence services of the second tier.

In principle, the \sm service deployment interface can verify the correctness
of an application before accepting it. In practice, this operation is very
hard to perform. One indicator of correctness is the compliance to a template
of acceptable interfaces for the kind of service the application provides.
However, it is very difficult to define templates strict enough to allow
sensible compliance checks, but general enough to avoid hindering the
deployment of legitimate services.

Another important detection strategy that we considered is to implement a
mechanism that could guarantee, in every moment, a reproducibility of the
results of a service. With provenance certifications of
raw data and their propagation to results, it is possible to implement a
reference monitor to verify compliance between results and expected values.
In case of conflicts between the declared results and the actual ones, \sm
could discover what has been tampered with: the source data, or the service
logic. This detection can also feed a data trustworthiness rating system.

Finally, another way to check correctness is to look at the actual behavior of
the application, as it is common in anti-malware checks. These
techniques are far from infallible, and their scope falls much shorter than
what is required in our setting. Indeed, in this context a malicious behavior
can be a subtle deviation from the correct
calculation~\cite{Malware_Analysis_limits}, which is far more difficult than
the detection of traditional malicious behaviors (e.g., damaging or
self-replicating ones). Promising techniques, which can benefit from the
execution of all the services on the \sm platform, are those based on the
aggregation of multi-domain information~\cite{6565228,althebyan2015mitigating}.


\subsubsection{Travel} 
\label{ssub:composition}
Services in the Travel category orchestrate Information ones to provide
highly coordinated functionalities to users. Since the services in this
category heavily rely on composition to provide their functionalities, their
main concerns regard their workflow.

\paragraph{Service Workflow Manipulation} Managers can modify the expected
flow of information among services for many purposes. As an example, consider
the Manager of a service called Bus ETA that predicts bus arrivals. In its
calculations, Bus ETA uses three source-services, respectively for traffic, GPS
positioning, and weather forecasts. Although the Manager preserves the logic
(i.e., the behavior) of the Bus ETA service, by simply changing the workflow,
i.e., the bindings of the Bus ETA to the other services, she can make (some) of
the sources unreachable, either completely disabling the Bus ETA service or
modifying the resulting output due to missing data.

\paragraph{Countermeasures} 
\sm already provide tools to contrast service workflow manipulations through
the helper services Dispatcher and Business Policies
(Fig~\ref{fig:small_architecture}). Indeed, when Managers deploy their
services in \sm, they also define the related access rules (stored and
retrieved in the Business Policies service). Then, all workflow compositions
pass through the Dispatcher service that logs them and enforces the established
access policies. In this way, unexpected workflows are detected, logged, and
(depending on the access rules) forbidden.
The monitoring capabilities of the Dispatcher can also be enhanced with
techniques like~\cite{schlicher2016towards}, where the authors propose an
analysis to detect malicious workflows and~\cite{dynamic}, that employs machine
learning engines similar to the ones used in dynamic malware analysis to
detect malicious workflows. Finally, based on service specifications, we can
create workflow graphs for strategic mitigation~\cite{velpula2009behavior}.

Another promising approach comes from the field of Choreographic
Programming~\cite{montesi2013choreographic}. The use of choreographies to
implement workflows among services is relatively
new~\cite{giallorenzo2016real}. We deem choreographies an effective prevention
tool that lets partners agree on a formal definition of their workflows, which
can be later compiled into their respective, compliant services. Moreover, in
the dynamic context of \sm, tools like~\cite{dalla2014aiocj} can aid partners
in updating their agreed workflows even at runtime (i.e., without stopping
their running services). These updates would be still conditioned to a general
agreement and maintain the same guarantees of the original services.

Mitigation techniques can be also developed following
e.g.,~\cite{goldberg2016explaining}. The idea would be to develop a \sm helper
service that monitors workflows and, once an attack by an insider is
discovered, it appropriately redirects the workflow to avoid further damage.

\paragraph{Composition of Unverified Services and Data} In the context of
mobility, verified information is of paramount importance. However, in a
service-oriented architecture, the tricky part to deal with is that a service
invocation can be seen as a collection of workflows. These workflows can
compose many levels of services, each processing and modifying the data before
its final destination.
These services inherently include the logic of the composed services and, by
extension, also the possible manipulations executed by insiders.
As an example, consider a journey planner that uses a real-time traffic report
service to avoid traffic jams and roadblocks. Since the journey planner
directly integrates the information from the traffic report service,
manipulating information of the latter alters the solutions of the journey
planner, diverting travelers towards certain pathways. This case presents an
interesting nuance: the insider is not a direct Manager of the considered
service (i.e., the journey planner), instead it is the Manager of a composed
service (the traffic report) that twists its contribution to alter the behavior
of the planner.
In this context also trustability, provenance, and trustworthiness of data
and/or services should be considered as possible targets of attacks. For
example, tampering with data provenance is a source of
attack~\cite{provenance_lit1} that in a MaaS scenario can see malicious
operators claiming to publish genuine data of a competitor, actually forging
them.

Interfering with the certification of data trustworthiness is another possible
vector. In this case, it is very difficult to block attacks in which, e.g.,
the creator advertises a data source of given quality, but then exposes a
degraded version to keep the advantage of more precise/timely information for
herself.
A related trustworthiness scenario is that of an insider who succeeds in
registering a rogue service. For example, a modified travel planner could
deflect routes to favor or damage certain businesses; a modified delay-checking
application could hide or amplify violations of agreed service levels.

\paragraph{Countermeasures}
A service must support the provision of different sources of data along with
their associated metadata (e.g., used to verify their provenance). However,
\sm shall also provide techniques, embodied by helper services, to transform
those data into verified information. Different approaches can be taken to
support a solution for the problem of recognizing the source of a data stream.
Literature agrees~\cite{groth2004protocol} that the requirements for a
provenance management system are:
\emph{Verifiability:} a provenance system should be able to verify a process
in terms of the actors (or services) involved, their actions, and their
relationship with one another;
\emph{Accountability}: an actor (or service) should be accountable for its
actions in a process. Thus, a provenance system should record in a
non-repudiable manner any provenance generated by a service;
\emph{Reproducibility}: a provenance system should be able to repeat a process
and possibly reproduce a process from the provenance stored;
\emph{Preservation}: a provenance system should have the ability to maintain
provenance information for an extended period of time. This is essential for
applications run in an enterprise system;
\emph{Scalability}: given the large amounts of data that an enterprise system
handles, a provenance system needs to be scalable;
\emph{Generality}: a provenance system should be able to record provenance
from a variety of applications;
\emph{Customizability}: a provenance system should allow users to customize it
by setting metadata such as time, events of recording, and the granularity of
provenance.

In these regards, it would be useful to deploy technologies to certify the
metadata related to a data stream and manage its validity during time and
re-elaboration~\cite{tsai2007data}. According to works
like~\cite{data_provenance_tech1}, this problem could be solved only with the
creation of a public-private key system for data stream certification. A good
reference is the system developed in~\cite{provenance_tech}, describing a
cryptographic provenance verification approach for ensuring data properties
and integrity for single hosts. Specifically, the authors designed and
implemented an efficient cryptographic protocol that enforces keystroke
integrity. This kind of protocols can be integrated as a helper service in
\sm. However, public-key schemes are known for their significant computational
load, thus existing techniques may not be suitable for high-rate, high-volume
data sources. Moreover, there could be the need for an algorithm for the
provenance of composed data.
In some cases, data originated from the composition of raw (or otherwise lower
ranked) sources should be accompanied by suitable metadata for verifying the
provenance of the input values, in a cryptographically strong way. In the
context of SMAll, it could be important and useful to capture and understand
the propagation of data.

The combination of metadata- with key-propagation management can guarantee a
good level of trust in provenance management systems. Works in the direction
of~\cite{he2015adding} discuss how to support provenance awareness in spatial
data infrastructure and investigates key issues including provenance modeling,
capturing, and sharing, useful to implement key propagation systems.

Finally, we address trustability, provenance, and trustworthiness of services and/or data.

Trustability needs to be measured by indicators for data quality and service
behavior. Values for these indicators come from a variety of considerations on
basic data sources. However, it is challenging to define algorithms for source
evaluation based on data resulting from services aggregating and orchestrating
other sources~\cite{data_quality1, data_quality2}.
Ascertaining provenance means ensuring that the source of data is verifiable,
i.e., that it corresponds to the one declared in the process of creation.
Trustworthiness is intended as the possibility to ascertain the
correctness of the information provided by a data source, which is loosely
related to provenance~\cite{data_trust_lit}. Ideally, but infrequently, data
samples can be independently measured by different users, thus allowing
cross-checking and error correction. For original data, i.e., provided by its
creator, the trustworthiness score is usually derived from the reputation of
the creator.
Clearly, guaranteeing data quality, provenance and trustworthiness is not
enough, it is necessary to ensure that the computation is correct and that no
useful results are hidden (completeness).


\subsubsection{User} 
\label{ssub:gateway}

The last category of services of tier I is not specific to mobility but it
contains essential functionalities for the other two categories. The most
representative case is that of User Profiling and Management. User
profiling is not required to create services for mobility, but it has become
essential to ensure usability, to provide user assistance, and to even
anticipate and plan for the next movements of the user (cf.~Google
Now\footnote{\url{https://www.google.com/search/about/learn-more/now/}}).

\paragraph{Data theft}
Here, the most obvious threat regards the possibility of stealing information derived
from the profile dataset, such as preferences, recordings of movements, orders
and payments. 


\paragraph{Countermeasures}
In our setting, a possible approach is to empower the user with control over
its profile and the related access policies~\cite{mydata}.




\subsection{MaaS Stack | Tier II} 
\label{sub:maas_stack_tier_ii}

\subsubsection{Business Intelligence} 
\label{ssub:business_intelligence}

The second tier of the MaaS Stack adds a new category next to the ones of the
first tier: Business Intelligence, i.e., services exclusively dedicated to
provide insight on the usage and performances of services of the first tier.

This services can implement any kind of data mining algorithm useful for
monitoring the profitability, sustainability, and reliability of the provided
services, as well as for determining trends and making predictions on future
usage, for capacity planning and policy definition.

\paragraph{Business Intelligence Data Theft}
Business Intelligence analyzes are important source of sensitive information
for insiders (also in this case Managers with privileged access) that could
expose relevant data to third parties. Indeed, without Business Intelligence
services it would be very difficult or even impossible for insiders to obtain
such data, that otherwise would require the access to massive amounts of
private information over long periods.

Managers of Business Intelligence services can apply targeted analyzes to
infer reserved information, such as policies and business strategies of their
company. An example of this type of attack is what we simulated
in~\cite{pajais}, where by just analyzing the database of validated 
tickets of a public transport company of the urban area of Bologna, 
we were able to
reconstruct the distribution of the various types of tickets in the different
zones of the city. 



\paragraph{Countermeasures}
\sm serves the purpose of mediating the access to relevant data for Business
Intelligence. Every operator wishing to obtain statistics or performance
indicators about its own services can freely create instances of the
platform-approved analytics services.

Regarding mitigation, the most effective way to hinder the possibility to
misuse Business Intelligence services is to properly sanitize the datasets and
to control the workflow of this information. These
techniques~\cite{mistry2015privacy} aim to prevent insiders from correlating
Business Intelligence services with external data sources to derive hidden
patterns or de-anonymize sensitive information.




\subsection{MaaS Stack | Tier III} 
\label{sub:maas_stack_tier_iii}

The third tier of the MaaS Stack is that of MaaS operators, i.e., eMobility
operators that use services of other companies, traded within a federated
market. In our case, \sm gives support to such a market but the creation of
dynamic federations of MaaS operators rises specific threats within \sm (and
MaaS markets in general).

In this scenario the main issues to consider are:
\begin{itemize}

	\item Data service management to avoid manipulation, impersonation, and
	sensitive pattern discovery (Prevention and Detection);

	\item Service workflow management to monitor invocation trends of services
	(Mitigation and Detection);

	\item Service quality and trustability management to verify the correctness
	of the service results (Prevention and Detection).

\end{itemize}

Indeed, the PaaS layer in \sm differs from most PaaS solutions. Traditionally
PaaS provides offer execution environments that isolate tenants. On the
contrary, \sm is built to ease the publication, integration, and orchestration
of services owned by different operators.

A simple example to clarify this characteristic is a one-stop ticketing
application that orchestrates:

\begin{itemize}
  \item a dynamic planner service providing routing options;
  \item a user profile manager to sort them according to user preferences;
  \item a real-time availability seat reservation service;
  \item a set of services for payment.
\end{itemize}

The hierarchy of the ticketing service spans many layers, e.g., it integrates
the dynamic planner that, in turns, orchestrates many services for static
(mapping, timetables) and real-time data (delays, planned extraordinary
events, disruptions). The composition of services forms a tree of dependencies
that reaches the level of raw-data information services, possibly branching
within the domains of different companies.

Since \sm aims at supporting this kind of interoperability, we argue that it
shall also assume responsibility for the trustworthiness and reliability
of the services; this is unusual for traditional PaaS~\cite{Not_as_Paas}.
Moreover, access control policies can be heterogeneous, exchanged data can
have different sensitivity levels, and the agents can be competing operators.

Clearly, the main insider threat for this scenario comes from the service
providers themselves, the MaaS operators. The malicious goals can be of various
kinds, spanning from the de-legitimization of services of competing operators,
to the theft of stored information such as policies or business
strategies, to insiders that apply mining techniques to infer these
information using the data available from their vantage point.

We now proceed by focusing our analysis on the relevant insider threats within
the categories of the third tier of the MaaS Stack.

\subsubsection{Roaming and Clearing} 
\label{ssub:clearing}
\sm aims at providing interoperability between different operators. In this
context, interoperability means that it is possible to implement ticketing
systems which seamlessly work on different operators across their zones of
influence. As mentioned in \S~\ref{sec:mobility_as_a_service}, this concept
(and the category of services that supports it) takes the name of Roaming.
Usually, to support at a business level the roaming for users among operators,
business agreements should be put into place to implement a Clearing system for
the redistribution of profits between transport operators. In this Section, we
consider threats as directed to the Clearing category since it comprises also
the threats to the Roaming one.

\paragraph{Pattern Extraction} As analyzed in
\cite{pajais}, the need for Clearing services is satisfied through a
centralized (federation-wise) system able to collect all the different data
sources from different operators and to perform economic evaluations. A
centralized clearing system scenario is typically based on a central database
that collects all the ticket validation data from every public transport
operator. This database is used both to perform economic evaluations to
redistribute profits and to store a permanent proof of the validity of this
evaluation. The clearing system must fulfill an effective trade-off between
public verifiability of the correctness of its operation and protection of
sensitive data provided by operators. As the last cited work shows, an insider
can perform data mining analysis and pattern discovery on the tickets datasets
in order to retrieve sensitive information about business strategies and
perform unfair competition.

\paragraph{Countermeasures}

\begin{figure}
\begin{center}
  \centering
  \includegraphics[width=\linewidth]{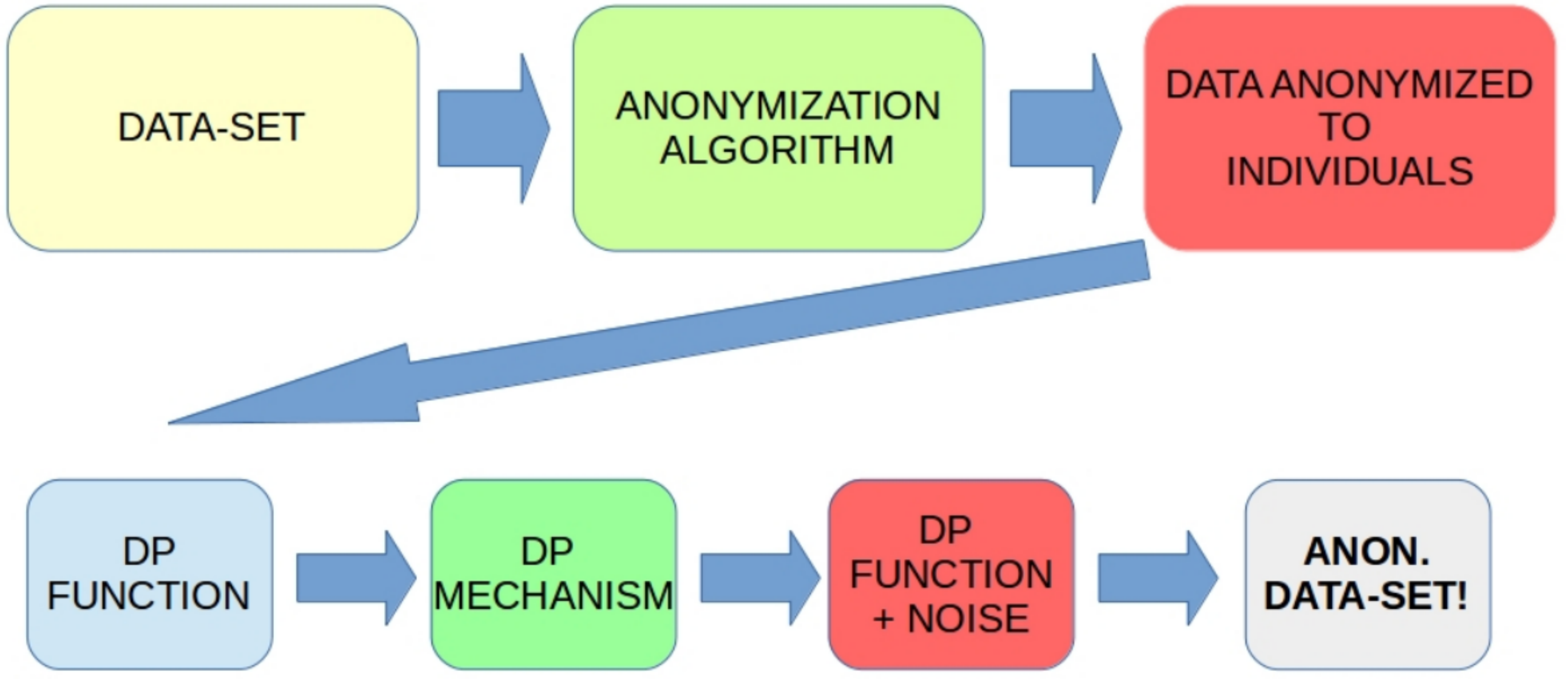}
  \caption{Data Sanitization algorithmic approach.}
  \label{fig:sanitization}
\end{center}
\end{figure}

To counteract \emph{Pattern Extraction}, it is possible to deploy sanitization
techniques~\cite{molnar2014automatic} able to mask the data enough to deny the
possibility to perform pattern analysis. These sanitization techniques balance
masking sensitive data and keeping enough properties and information required
to perform the economic evaluations. In order to do what we described, we could
assemble an anonymization system, as shown in Fig.~\ref{fig:sanitization}, that
combines masking techniques for the raw dataset (once deployed in the
centralized database clearing system) and a differential privacy engine able to
introduce a certain amount of noise and prevent exploit techniques as cross-combining data with external ones.

\subsubsection{Access Control and Service Level Agreement}
\label{ssub:access_control_and_service_level_agreement}
Service Level Agreement (SLA) and Access Control (AC) services in \sm are meant
to throttle the invocation of tier I services provided by an operator on the
basis of commercial agreements with other operators. It is possible to see SLA
as a contract ruling the quantity or rate of invocation of each service, and AC
as a contract ruling the quality or the set of provided data or services.
Obviously, malicious insiders may try to circumvent these limitations.

\paragraph{Countermeasures}
When a SLA or an AC policy is in place, all service invocations must be tracked
(or even proxied) by an infrastructural service provided by \sm. This makes
evading enforcement difficult. The most common vulnerability in this context is
not tied to policy enforcement, however, but rather to policy specification. To
this end, \sm could restrict acceptable policies to those drafted with an
internal helper service, following a standard framework, and formally verifying
their soundness before applying them. Access control models and formal policy
specification languages have been around for some
time~\cite{sandhu1996role,Ponder}, and they have evolved into sophisticated,
standardized models like ABAC~\cite{hu2013guide,sandhu2015attribute}.
Inadequate (but consistent) policy definitions due to poor understanding of the
federation interactions or to carelessness cannot be tackled at this level;
logging and auditing facilities integrated in \sm provide valuable feedback at
run-time about the effectiveness of installed policies.


\subsubsection{Business Intelligence} 
\label{ssub:business_intelligence}

Similarly to tier II, in tier III we have a category of services dedicated to
business intelligence. The difference with respect to the services of the
second tier is that here the analyzes span data belonging to a multitude of
operators. Indeed, as it happens for clearing services, the business
intelligence services of the third tier relate to the management of data,
statistics, and administration of services shared among operators. The
availability of such aggregated data can give free access to companies (seen as
federated insiders) to data and analyzes of competitors.
Referring again the case of the dynamic route planner as a running example, the
service can use real-time data of different companies to take into account the
average delays of transport vehicles in the calculation of its solutions. The
averaged delays are the result of a business intelligence service that collects
all the delays of a route within a specific area that involves several
operators and calculates the delays. Finally, the recorded delays are collected
into a shared dataset accessible by all the participants.

In this example, an insider can use the collected dataset to find out where the
competitors operate with bigger delays and profit from this information by
exposing their faults to the regional administration.
Insiders can also expose cartels where operators systematically provide a bad
service during rush hours to favor a specific company (e.g., because they hold
some economic interest in it). Finally, the insiders can also find out if an
operator hides delays making analysis on the correspondent road conditions
(e.g., showing that buses could not sustain certain speeds since their routes
were jammed).

\paragraph{Countermeasures} 
All the countermeasures for this kind of attacks are based on a trade-off
between the amount of sensitive data preserved and utility of the queries.
Different anonymization and sanitization techniques have been proposed for
complex datasets, but since in \sm Business Intelligence services share the
results of queries, we need to introduce a measure that indicates the maximum
amount of anonymized information such that the queries still work.

Different works proposed metrics for the evaluation of the amount of privacy
preserved in specific dataset. A measure introduced in~\cite{nergiz2007hiding}
defined an evaluation metric about the presence of pattern in a dataset called
$\delta$-presence. We can use this metric to evaluate the presence of a
specific patterns in the shared dataset. Another interesting work in this
direction is~\cite{heydt2006privacy} which operates by complementing existing
techniques with post randomization methods.






\section{Conclusions}
\label{sec:conclusions}
In this paper, we presented the concept of Mobility as a Service and how MaaS
operators shall facilitate the dynamic provisioning of multi-modal
transportation to their users. To support such flexibility we are developing a
federated marketplace of services called \sm, aimed at harmonizing data
flows and service invocations.

This kind of federated platform is particularly sensitive to insider threats,
which emerge at different layers, targeting both the constituent components
provided by users and operators and the services provided by
the platform itself.

The MaaS Stack, our tiered view on the components of MaaS markets, allowed us
to treat in isolation the security issues of each tier. Often, these issues
turn out to be instances of well-known threats in the fields of cloud
computing, service-oriented architectures, supply chain management, and
trusted business partnerships.

In principle, the platform allows to implement context-specific versions of the
solutions proposed in the literature regarding the aforementioned fields, as
well as novel solutions inspired by their cross-fertilization. We argue that
the central role of \sm in mediating every interaction and in collecting their
traces makes the platform fit to host solutions to the presented security
issues of MaaS markets.

The effectiveness of the proposed approaches will be experimentally validated
in the near future, following the completion of the platform in all its parts,
and the deployment of real-world services on it.

\bibliographystyle{ieeetr}
\bibliography{biblio}

\begin{thebibliography}{10}

\bibitem{Buyya2009599}
R.~Buyya, C.~S. Yeo, S.~Venugopal, J.~Broberg, and I.~Brandic, ``Cloud
  computing and emerging {IT} platforms: Vision, hype, and reality for
  delivering computing as the 5th utility,'' {\em FGS}, vol.~25, no.~6, pp.~599
  -- 616, 2009.

\bibitem{MaaS}
S.~Pippuri, S.~Hietanen, and K.~Pyyhtiä, ``Maas finland.''
  \url{http://maas.fi/}.

\bibitem{rochwerger2009reservoir}
B.~Rochwerger, D.~Breitgand, E.~Levy, A.~Galis, K.~Nagin, I.~M. Llorente,
  R.~Montero, Y.~Wolfsthal, E.~Elmroth, J.~Caceres, {\em et~al.}, ``The
  reservoir model and architecture for open federated cloud computing,'' {\em
  IBM JRD}, vol.~53, no.~4, pp.~4--1, 2009.

\bibitem{buyya2010intercloud}
R.~Buyya, R.~Ranjan, and R.~N. Calheiros, ``Intercloud: Utility-oriented
  federation of cloud computing environments for scaling of application
  services,'' in {\em AAPP}, pp.~13--31, Springer, 2010.

\bibitem{Kandias2013}
M.~Kandias, N.~Virvilis, and D.~Gritzalis, {\em The Insider Threat in Cloud
  Computing}, pp.~93--103.
\newblock Springer, 2013.

\bibitem{zhou2012long}
P.~Zhou, Y.~Zheng, and M.~Li, ``How long to wait?: predicting bus arrival time
  with mobile phone based participatory sensing,'' in {\em Proceedings of
  MobiSys}, pp.~379--392, ACM, 2012.

\bibitem{mirri2016}
S.~Mirri, A.~Melis, C.~Prandi, and M.~Prandini, ``Crowdsensing for smart
  mobility through a service-oriented architecture,'' in {\em ISCC}, p.~5,
  IEEE, 2016.

\bibitem{maaS_stack}
S.~Giallorenzo, A.~Melis, and M.~Prandini, ``{S}mart {M}obility for {A}ll,''
  tech. rep., University of Bologna, 2016.

\bibitem{mouly1992gsm}
M.~Mouly, M.-B. Pautet, and T.~Foreword By-Haug, {\em The GSM system for mobile
  communications}.
\newblock Telecom publishing, 1992.

\bibitem{mun2008yet}
H.~Mun, K.~Han, C.~Yeun, and K.~Kim, ``Yet another intrusion detection system
  against insider attacks,'' {\em Proc. of SCIS}, 2008.

\bibitem{Stavrou2014}
V.~Stavrou, M.~Kandias, G.~Karoulas, and D.~Gritzalis, {\em Business Process
  Modeling for Insider Threat Monitoring and Handling}, pp.~119--131.
\newblock Cham: Springer International Publishing, 2014.

\bibitem{flynn2014cloud}
L.~Flynn, G.~Porter, and C.~DiFatta, ``Cloud service provider methods for
  managing insider threats: Analysis phase ii, expanded analysis and
  recommendations,'' 2014.

\bibitem{claycomb2012insider}
W.~R. Claycomb and A.~Nicoll, ``Insider threats to cloud computing: Directions
  for new research challenges,'' in {\em ACSAC}, pp.~387--394, IEEE, 2012.

\bibitem{schneier2011secrets}
B.~Schneier, {\em Secrets and lies: digital security in a networked world}.
\newblock John Wiley \& Sons, 2011.

\bibitem{nostro2014insider}
N.~Nostro, A.~Ceccarelli, A.~Bondavalli, and F.~Brancati, ``Insider threat
  assessment: A model-based methodology,'' {\em ACM SIGOPS}, vol.~48, no.~2,
  pp.~3--12, 2014.

\bibitem{casey}
T.~Casey, ``A field guide to insider threat,'' tech. rep., Intel, 2015.

\bibitem{silowash2012common}
G.~Silowash, D.~Cappelli, A.~Moore, R.~Trzeciak, T.~J. Shimeall, and L.~Flynn,
  ``Common sense guide to mitigating insider threats 4th edition,'' tech. rep.,
  DTIC Document, 2012.

\bibitem{bishop2005position}
M.~Bishop, ``Position: Insider is relative,'' in {\em Proceedings of Workshop
  on New security paradigms}, pp.~77--78, ACM, 2005.

\bibitem{hunker2011insiders}
J.~Hunker and C.~W. Probst, ``Insiders and insider threats-an overview of
  definitions and mitigation techniques.,'' {\em JoWUA}, vol.~2, no.~1,
  pp.~4--27, 2011.

\bibitem{shabtai2012survey}
A.~Shabtai, Y.~Elovici, and L.~Rokach, {\em A survey of data leakage detection
  and prevention solutions}.
\newblock Springer, 2012.

\bibitem{mirri3_2016}
S.~Mirri, A.~Melis, C.~Prandi, and M.~Prandini, ``A microservice architecture
  use case for persons with disabilities,'' in {\em CVSJ}, p.~5, Hindawi, 2016.

\bibitem{insider}
Y.~Cho, G.~Qu, and Y.~Wu, ``Insider threats against trust mechanism with
  watchdog and defending approaches in wireless sensor networks,'' in {\em
  SPW}, pp.~134--141, IEEE, 2012.

\bibitem{gertz2007handbook}
M.~Gertz and S.~Jajodia, {\em Handbook of database security: applications and
  trends}.
\newblock Springer, 2007.

\bibitem{shatnawi2011detection}
N.~Shatnawi, Q.~Althebyan, and W.~Mardini, ``Detection of insiders misuse in
  database systems,'' in {\em ICECS}, vol.~1, 2011.

\bibitem{spitzner2003honeypots}
L.~Spitzner, ``Honeypots: Catching the insider threat,'' in {\em CSAC},
  pp.~170--179, IEEE, 2003.

\bibitem{mirri2_2016}
S.~Mirri, A.~Melis, C.~Prandi, and M.~Prandini, ``A service-oriented approach
  to crowdsensing for accessible smart mobility scenarios,'' in {\em
  Proceedings ICCTS}, p.~5, IEEE, 2016.

\bibitem{ho2016demystifying}
S.~M. Ho, J.~T. Hancock, C.~Booth, M.~Burmester, X.~Liu, and S.~S. Timmarajus,
  ``Demystifying insider threat: Language-action cues in group dynamics,'' in
  {\em HICSS}, pp.~2729--2738, IEEE, 2016.

\bibitem{althebyan2008design}
Q.~Althebyan, {\em Design and analysis of knowledge-base centric insider threat
  models}.
\newblock ProQuest, 2008.

\bibitem{kamra2008detecting}
A.~Kamra, E.~Terzi, and E.~Bertino, ``Detecting anomalous access patterns in
  relational databases,'' {\em The VLDB Journal}, vol.~17, no.~5,
  pp.~1063--1077, 2008.

\bibitem{mathew2010data}
S.~Mathew, M.~Petropoulos, H.~Q. Ngo, and S.~Upadhyaya, ``A data-centric
  approach to insider attack detection in database systems,'' in {\em RAID},
  pp.~382--401, Springer, 2010.

\bibitem{doss2009developing}
G.~Doss and G.~Tejay, ``Developing insider attack detection model: a grounded
  approach,'' in {\em ISI}, pp.~107--112, IEEE, 2009.

\bibitem{Malware_Analysis_limits}
A.~Moser, C.~Kruegel, and E.~Kirda, ``Limits of static analysis for malware
  detection,'' in {\em ACSAC}, pp.~421--430, Dec 2007.

\bibitem{6565228}
H.~Eldardiry, E.~Bart, J.~Liu, J.~Hanley, B.~Price, and O.~Brdiczka,
  ``Multi-domain information fusion for insider threat detection,'' in {\em
  SPW}, pp.~45--51, May 2013.

\bibitem{althebyan2015mitigating}
Q.~Althebyan, R.~Mohawesh, Q.~Yaseen, and Y.~Jararweh, ``Mitigating insider
  threats in a cloud using a knowledgebase approach while maintaining data
  availability,'' in {\em ICITST}, pp.~226--231, IEEE, 2015.

\bibitem{schlicher2016towards}
B.~G. Schlicher, L.~P. MacIntyre, and R.~K. Abercrombie, ``Towards reducing the
  data exfiltration surface for the insider threat,'' in {\em HICSS},
  pp.~2749--2758, IEEE, 2016.

\bibitem{dynamic}
M.~D. Ernst, ``Static and dynamic analysis: Synergy and duality,'' in {\em
  WODA}, pp.~24--27, Citeseer, 2003.

\bibitem{velpula2009behavior}
V.~B. Velpula and D.~Gudipudi, ``Behavior-anomaly-based system for detecting
  insider attacks and data mining,'' {\em IJRTE}, vol.~1, no.~2, pp.~261--266,
  2009.

\bibitem{montesi2013choreographic}
F.~Montesi, {\em Choreographic Programming}.
\newblock PhD thesis, IT University of Copenhagen, 2013.

\bibitem{giallorenzo2016real}
S.~Giallorenzo, {\em Real-World Choreographies}.
\newblock PhD thesis, Universit{\`a} degli studi di Bologna, 2016.

\bibitem{dalla2014aiocj}
M.~Dalla~Preda, S.~Giallorenzo, I.~Lanese, J.~Mauro, and M.~Gabbrielli,
  ``{AIOCJ}: A choreographic framework for safe adaptive distributed
  applications,'' in {\em SLE}, pp.~161--170, Springer, 2014.

\bibitem{goldberg2016explaining}
H.~G. Goldberg, W.~T. Young, A.~Memory, and T.~E. Senator, ``Explaining and
  aggregating anomalies to detect insider threats,'' in {\em HICSS},
  pp.~2739--2748, IEEE, 2016.

\bibitem{provenance_lit1}
Y.~L. Simmhan, B.~Plale, and D.~Gannon, ``A survey of data provenance in
  e-science,'' {\em SIGMOD Rec.}, vol.~34, pp.~31--36, Sept. 2005.

\bibitem{groth2004protocol}
P.~Groth, M.~Luck, and L.~Moreau, ``A protocol for recording provenance in
  service-oriented grids,'' in {\em PDS}, pp.~124--139, Springer, 2004.

\bibitem{tsai2007data}
W.-T. Tsai, X.~Wei, Y.~Chen, R.~Paul, J.-Y. Chung, and D.~Zhang, ``Data
  provenance in {SOA}: security, reliability, and integrity,'' {\em SOCA},
  vol.~1, no.~4, pp.~223--247, 2007.

\bibitem{data_provenance_tech1}
Y.~L. Simmhan, B.~Plale, and D.~Gannon, ``A survey of data provenance
  techniques,'' {\em CSD, IU, Indiana}, vol.~47405, 2005.

\bibitem{provenance_tech}
K.~Xu, H.~Xiong, C.~Wu, D.~Stefan, and D.~Yao, ``Data-provenance verification
  for secure hosts,'' {\em DSC}, vol.~9, pp.~173--183, March 2012.

\bibitem{he2015adding}
L.~He, P.~Yue, L.~Di, M.~Zhang, and L.~Hu, ``Adding geospatial data provenance
  into {SDI} — a service-oriented approach,'' {\em AEORS}, vol.~8, no.~2,
  pp.~926--936, 2015.

\bibitem{data_quality1}
C.~Falge, B.~Otto, and H.~Österle, ``Data quality requirements of
  collaborative business processes,'' in {\em HICSS}, pp.~4316--4325, Jan 2012.

\bibitem{data_quality2}
S.~Dustdar, R.~Pichler, V.~Savenkov, and H.-L. Truong, ``Quality-aware
  service-oriented data integration: Requirements, state of the art and open
  challenges,'' {\em SIGMOD Rec.}, vol.~41, pp.~11--19, Apr. 2012.

\bibitem{data_trust_lit}
C.~Dai, D.~Lin, E.~Bertino, and M.~Kantarcioglu, {\em SDM}, ch.~An Approach to
  Evaluate Data Trustworthiness Based on Data Provenance, pp.~82--98.
\newblock Berlin, Heidelberg: Springer, 2008.

\bibitem{mydata}
K.~K. Antti~Poikola and H.~Honko, ``Mydata a nordic model for human-centered
  personal data management and processing,'' tech. rep., Ministry of Transport
  Finland, 2014.

\bibitem{pajais}
F.~Callegati, A.~Campi, A.~Melis, M.~Prandini, and B.~Zevenbergen,
  ``Privacy-preserving design of data processing systems in the public
  transport context,'' {\em Pacific Asia Journal of the Association for
  Information Systems}, vol.~7, no.~4, 2015.

\bibitem{mistry2015privacy}
B.~R. Mistry and A.~Desai, ``Privacy preserving heuristic approach for
  association rule mining in distributed database,'' in {\em ICIIECS},
  pp.~1--7, IEEE, 2015.

\bibitem{Not_as_Paas}
S.~E. Madnick, R.~Y. Wang, Y.~W. Lee, and H.~Zhu, ``Overview and framework for
  data and information quality research,'' {\em J. Data and Information
  Quality}, vol.~1, pp.~2:1--2:22, June 2009.

\bibitem{molnar2014automatic}
D.~Molnar, B.~Livshits, P.~Godefroid, and P.~Saxena, ``Automatic
  context-sensitive sanitization,'' Nov.~25 2014.
\newblock US Patent 8,898,776.

\bibitem{sandhu1996role}
R.~S. Sandhu, E.~J. Coynek, H.~L. Feinsteink, and C.~E. Youmank, ``Role-based
  access control models yz,'' {\em IEEE computer}, vol.~29, no.~2, pp.~38--47,
  1996.

\bibitem{Ponder}
N.~Damianou, N.~Dulay, E.~Lupu, and M.~Sloman, ``The ponder policy
  specification language,'' in {\em IWPDSN}, POLICY '01, (London, UK, UK),
  pp.~18--38, Springer-Verlag, 2001.

\bibitem{hu2013guide}
V.~C. Hu, D.~Ferraiolo, R.~Kuhn, A.~R. Friedman, A.~J. Lang, M.~M. Cogdell,
  A.~Schnitzer, K.~Sandlin, R.~Miller, K.~Scarfone, {\em et~al.}, ``Guide to
  attribute based access control (abac) definition and considerations
  (draft),'' {\em NIST Special Publication}, vol.~800, no.~162, 2013.

\bibitem{sandhu2015attribute}
R.~Sandhu, ``Attribute-based access control models and beyond.,'' in {\em
  ASIACCS}, p.~677, 2015.

\bibitem{nergiz2007hiding}
M.~E. Nergiz, M.~Atzori, and C.~Clifton, ``Hiding the presence of individuals
  from shared databases,'' in {\em SIGMOD}, pp.~665--676, ACM, 2007.

\bibitem{heydt2006privacy}
T.~S. Heydt-Benjamin, H.-J. Chae, B.~Defend, and K.~Fu, ``Privacy for public
  transportation,'' in {\em IWPET}, pp.~1--19, Springer, 2006.

\end{thebibliography}



\end{document}